\documentclass[12pt,twoside]{article}
\begin{document}
\pagestyle{myheadings} \markboth{Second Quantization and
Bogoliubov Approximation}{Bayram Kurucay}

\title{Second Quantization and Bogoliubov Approximation}
\author{Bayram Kurucay}
\date{08.July.2006}
\maketitle

\begin{abstract}
Recent experiments with trapped alkali atoms
[\ref{ref:Experiment}] have drawn enormous interest to the
theoretical studies concerning Bose-Einstein condensation. The
purpose of this paper is to review one of the approaches to study
bosonic matter at zero temperature, namely the Bogoliubov
approximation. Review of a necessary tool, the second
quantization, will also be made.
\end{abstract}

\section{Introduction}
In 1925, Albert Einstein, by generalizing Satyendra Nath Bose's
work [\ref{ref:Bose}] on photons, concluded [\ref{ref:Einstein}]
that macroscopic numbers of integer spinned atoms can collapse
into one single quantum state if they are cooled to temperatures
very close to the absolute zero. However, because of technical
limitations of trapping technology, this phenomenon, known as
Bose-Einstein Condensation (BEC), has not been observed until
1995, when more than one research group almost simultaneously
managed to obtain BECs using trapped alkali atoms
[\ref{ref:Experiment}].

A common point of all these experiments is that they worked with
weakly interacting atoms. The Bogoliubov approximation, which has
long been known to theorists, provides a proper means to approach
such systems. Although it is completely valid only at absolute
zero temperature, it still gives admissible results in the
temperature ranges concerned in the experiments.

The purpose of this paper is to review the Bogoliubov
approximation. An essential tool to accomplish this is the
\emph{second quantization}, which is an extension of ordinary
quantum mechanics. In part 2, it will be thoroughly studied.
Bogoliubov approximation will be discussed in the third part.

\section{Second Quantization}
Second quantization is a powerful tool. First introduced by
Jordan, Klein and Wigner in 1928 [\ref{ref:2nd}], it is now
covered in many text books
[\ref{ref:Fetter_Walecka}][\ref{ref:Greiner}]. It can be used for
field quantization, which makes it more powerful than the
'ordinary' quantum mechanics, that is the 'first quantization'. It
can also be used to study the properties of many particle systems.
Although it does not say anything more than the first quantization
for that purpose, it introduces convenience. This is the reason
that we will use it to approach Bose-Einstein Condensation.

Let us consider a specific and useful example rather than
considering the most general case. Harmonic potentials are very
common in physics and therefore familiar to  physicists.
Furthermore, to first approximation, every potential is harmonic
around its equilibrium point. Also, in 1D there is no degeneracy
and energy levels take the simple form $E_n=\hbar
\omega(n+\frac{1}{2})$. Therefore, we choose to study a harmonic
potential. The Hamiltonian of a system of N indistinguishable
particles confined in a 1D harmonic trap is given as
\begin{equation}
H=\sum_{k=1}^{N}{(-\frac{\hbar^2}{2m} \nabla_k^2+m\omega^2
r_{k}^2)} +\frac{1}{2}\sum_{k,l=1 ; k \neq l }^{N}{V(r_{k}-r_{l})}
.\end{equation}

We will assume an interparticle potential of the
form
\begin{equation}
V(r_{k}-r_{l})=g \delta(r_{k}-r_{l}) .
\end{equation}

In general, the N-body wave function $\Psi(r_1,r_2,...,r_N,t)$ can
be written as a superposition of products of one body stationary
wave functions $\phi_{k}(r_i)$:
\begin{equation}
\Psi(r_1,r_2,...,r_N,t)=\sum_{k_1,k_2,...,k_N}{C(k_1,k_2,...,k_N,t)\cdot\phi_{k_1}(r_1)\cdot\phi_{k_2}(r_2)...\cdot\phi_{k_N}(r_N)}
\end{equation}

According to postulates of quantum mechanics, all existing
particles can be divided into two subgroups: \emph{fermions},
which are the particles with half-integer spin, and \emph{bosons},
which are the ones with integer spin. The wave function of a
system consisting of identical bosons must be symmetric with
respect to the exchange of two particles, whereas that of a system
consisting of fermions must be anti-symmetric. Mathematically
stated, this implies
\begin{equation}
\Psi(...,r_i,...,r_j,...,t)=\Psi(...,r_j,...,r_i,...,t)
\end{equation}
 for bosons and
 \begin{equation}
\Psi(...,r_i,...,r_j,...,t)=-\Psi(...,r_j,...,r_i,...,t)
\end{equation}
for fermions. Since Bose-Einstein condensation occurs only in
 systems of identical bosons, fermionic systems will henceforth be
 omitted from this paper and unless otherwise stated, the equations
 will be valid only for bosons.

This is a natural point to raise the question "How many different
states may exist with $n_1$ particles having the lowest energy
$E_1$, $n_2$ particles having $E_2$, etc.?" It is not very hard to
see (yet not obvious either) that for bosons (and also for
fermions) the symmetry (anti-symmetry) condition precludes all but
one such state for a given set of occupation numbers
${n_1,n_2,...,n_N}$. That is to say, the occupation numbers
${n_i}$ determine the wave function uniquely. This fact is due to
the indistinguishability of the particles. In fact, if there could
be more than one state with the same occupation numbers, this
would be in contradiction with the indistinguishability.

As the occupation numbers define the wave function uniquely, it is
possible to span the space by vectors defined by the occupation
numbers. The space spanned by states of definite occupation
numbers is called the Fock space. A typical vector in the Fock
space is denoted by $|n_1 n_2 ... n_i...\rangle$ which means that
there are $n_1$ particles with $E_1$ etc.. Note that here $i$ can
be any integer regardless of the total particle number, so long as
there is no restriction in the highest energy that a particle can
have, even though there are a finite number of particles in the
system.

Annihilation and creation operators for each of the energy
eigenstates can be defined similarly to those encountered in the
harmonic oscillator problems. $\hat{b}_i^\dagger$ is defined as
the operator that creates a particle in the $E_i$ energy
eigenstate, that is,
\begin{equation}
\hat{b}_i^\dagger |n_1 n_2 ... n_i...\rangle=\sqrt{n_i+1}|n_1 n_2
... (n_i+1)...\rangle .
\end{equation}

The algebra of these operators is similar to that of harmonic
oscillator operators, with some additions to define commutators of
operators belonging to different modes:
\begin{equation}
\hat{b}_i |n_1 n_2 ... n_i...\rangle=\sqrt{n_i}|n_1 n_2 ...
(n_i-1)...\rangle
\end{equation}
\begin{equation}
[\hat{b}_i,\hat{b}_j]=0
\end{equation}
\begin{equation}
[\hat{b}_i^\dagger,\hat{b}_j^\dagger]=0
\end{equation}
\begin{equation}
[\hat{b}_i,\hat{b}_j^\dagger]=\delta_{ij}
\end{equation}

The relationship between second quantization and ordinary quantum
mechanics can be established once we define the \emph{field
operators}:
\begin{equation}
\hat{\psi}(r,t)=\sum_{n}{b_n \phi_n (r)}
\end{equation}
\begin{equation}
\hat{\psi}^\dagger (r,t)=\sum_{n}{b_n^\dagger  \phi_n^* (r)}
\end{equation}

It is not hard to prove that given the orthonormality of the basis
states, the field operator $\hat{\psi}(x,t)$ acting on the vacuum
state $|0,0,...,0,...\rangle$ creates a particle at position $x$
at time $t$. (At this point, the reader is urged not to confuse
the vacuum state, which has no particles at all, with
$|N,0,...,0,...\rangle$, the ground state of a system of N
particles.) Similarly, the expression $\int{f(x) \hat{\psi}(x,t)
dx}$ creates a particle with wave function $f(x)$. (Note that as
$\int{\phi_{k}(x) \hat{\psi}(x,t) dx}$ creates a particle in the
$k^{th}$ eigenstate, it can be proven to be equal to $\hat{b}_k$.)

The commutators of field operators are also important:
\begin{equation}
[\hat{\psi}(r,t),\hat{\psi}(r',t)]=0
\end{equation}
\begin{equation}
[\hat{\psi}^\dagger (r,t),\hat{\psi}^\dagger (r',t)]=0
\end{equation}
\begin{equation}
[\hat{\psi}(r,t),\hat{\psi}^\dagger (r',t)]=\delta_{r r'}
\end{equation}

The second quantized Hamiltonian $\hat{H}$ follows from the
expectation value of H with the field operators. The one particle
part of it becomes:
\begin{equation}
\hat{H}_0=\int{dx \hat{\psi}^\dagger(x,t)[-\frac{\hbar^2}{2m}
\nabla^2+m\omega^2 x^2] \hat{\psi}(x,t)} .
\end{equation}

Some tedious but straightforward algebra is sufficient to show
that
\begin{equation}
\hat{H}_0=\sum_n {E_n \hat{b}^\dagger_n \hat{b}_n} .
\end{equation}
Two particle operators are more tedious to handle, since their
second quantized forms are given [\ref{ref:Fetter_Walecka}] by
\begin{equation}
\hat{V}=\frac{1}{2}\int\int dr dr' \hat{\psi}^\dagger
(r)\hat{\psi}^\dagger (r') V(r,r')\hat{\psi}(r') \hat{\psi}(r')
\end{equation}

However, the the assumed Dirac delta form of the interparticle
potential allows us to reduce the double integral into a single
one:
\begin{equation}
\hat{V}=\frac{g}{2}\int dr \hat{\psi}^\dagger
(r)\hat{\psi}^\dagger (r') \hat{\psi}(r') \hat{\psi}(r')
\end{equation}

Although what is covered here is only a small part of the second
quantization, these expressions give us enough facilities to
consider the Bogoliubov approximation.

\section{Bogoliubov Approximation}

Bose-Einstein condensation occurs when the macroscoping majority
of particles fall into the ground state, that is when the number
of particles in the ground state, $n_0$, is $n_0 \approx N$ and
therefore, $n_{i}/N \ll 1$ for $i \neq 0$. Therefore, it may prove
to be convenient to write the field operator as
\begin{equation}
\hat{\psi}(r)=\phi_0 (r) \hat{b}_0 + \sum_{i=1}^{\infty}{\phi_i
(r) \hat{b}_i}.
\end{equation}

Considering the fact that $|\hat{b}_i |...n_{i}...\rangle | \sim
n_i$, the second term in this expression may be considered to be a
small fluctuation:
\begin{equation}
\hat{\psi}(r)=\phi_0 (r) \hat{b}_0 + \delta \hat{\psi} (r,t).
\end{equation}

The validity of this approach can be questioned, since depending
on the temperature the fluctuation part may become large and
cannot be assumed to be small anymore. However, for Bose-Einstein
condensation to occur, the temperature ranges considered should be
very close to the absolute zero; hence, this approach is quite
valid - though not flawless.

Now let us see the effect of the ground state annihilation
operator $\hat{b}_0$ on the Bose-Einstein condensed state:
\begin{equation}
\hat{b}_0 |n_0 ,...\rangle=\sqrt{n_0}|n_{0}-1,...\rangle .
\end{equation}

But as we talk of \emph{macroscopic} number of bosons in the
ground state, the difference between $n_0$ and $n_0-1$ is
negligible in the sense that $(n_0-1)-n_0 \simeq 1$ and hence we
can assume
\begin{equation}
\hat{b}_0 |n_0 ,...\rangle=\sqrt{n_0}|n_{0},...\rangle .
\end{equation}

So, the sole effect of the ground state annihilation operator is
to multiply the state with the real number $n_0$. The idea behind
Bogoliubov's approximation is to replace this operator $\hat{b}$
(and also its conjugate $\hat{b}^\dagger$) with the real number
$\sqrt{n_0}$:
\begin{equation}
\hat{b}_0\rightarrow\sqrt{n_0}.
\end{equation}
\begin{equation}
\hat{b}_0^\dagger \rightarrow\sqrt{n_0}.
\end{equation}

Hence, we obtain a field operator which differs from the
stationary ground state wave function only by a small fluctuation
part:
\begin{equation}
\hat{\psi}(r)=\sqrt{n_0} \phi_0 (r) + \delta \hat{\psi} (r,t).
\end{equation}

This completes our review. The second quantized Hamiltonian will
be much easier to calculate when the field operator is used in
this format, as the commutators will take simpler forms. Since
$\delta \hat{\psi} (r,t)$ is assumed to be small, terms which are
higher order with respect to it may be neglected. Omitting
$3^{rd}$ and $4^{th}$ orders and choosing a ground state wave
function which makes $1^{st}$ order terms vanish is a common
approach [\ref{ref:Fetter72}].

\newpage
\appendix
\section{References}
\begin{enumerate}
\item{\label{ref:Bose}S. N. Bose, Z. Phys. \textbf{26}, 178
(1924).}

\item{\label{ref:Einstein} A. Einstein, Sitz. Ber. Preuss. Akad.
Wiss. (Berlin) \textbf{1}, 3 (1925).}

\item{\label{ref:Experiment} Anderson \emph{et al.}, Science
\textbf{269}, 198 (1995); Bradley \emph{et al.}, Phys. Rev. Lett.
\textbf{75}, 1687 (1995); Davis \emph{et al.}, Phys. Rev. Lett.
\textbf{75}, 3969 (1995).}

\item{\label{ref:2nd}Jordan, Klein, Wigner, Z. Phys. \textbf{47},
631(1928).}

\item{\label{ref:Fetter_Walecka}Fetter, Walecka, \emph{Quantum
Theory of Many Particle Systems}, (McGraw-Hill, 1971).}

\item{\label{ref:Greiner} Greiner, \emph{Quantum Mechanics -
Special Chapters}, (Springer Verlag, 1998).}

\item{\label{ref:Fetter72}Fetter, Ann. Phys. \textbf{70}, 67
(1972).}

\end{enumerate}

\end{document}